\documentclass[pra,twocolumn,showpacs,amsmath,amssymb,superscriptaddress]{revtex4-1}

\usepackage{color}
\usepackage{amssymb}
\usepackage{amsthm}
\usepackage{graphicx}
\usepackage{epsfig}
\usepackage{subfigure}
\usepackage{amsmath,amssymb}
\usepackage[colorlinks=true,citecolor=blue,urlcolor=blue]{hyperref}
\usepackage{color}

\newcommand{\be}{\begin{equation}}
\newcommand{\ee}{\end{equation}}
\newcommand{\ba}{\begin{eqnarray}}
\newcommand{\ea}{\end{eqnarray}}
\newcommand{\ban}{\begin{eqnarray*}}
\newcommand{\ean}{\end{eqnarray*}}

\newcommand{\ket}[1]{\mbox{$ | #1 \rangle $}}
\newcommand{\bra}[1]{\mbox{$ \langle #1 | $}}

\newcommand{\one}{\leavevmode\hbox{\small1\normalsize\kern-.33em1}}

\begin{document}

\title{\Large\textbf{Breakdown of local convertibility through Majorana modes \\in a quantum quench}}
\author{Li Dai}
\affiliation{Department of Physics, National Chung Hsing University, Taichung, 40227, Taiwan}

\author{Ming-Chiang Chung}
\email{mingchiangha@phys.nchu.edu.tw}
\affiliation{Department of Physics, National Chung Hsing University, Taichung, 40227, Taiwan}
\affiliation{Physics Division, National Center for Theoretical Sciences, Hsinchu, 30013, Taiwan}

\begin{abstract}
The local convertibility of quantum states, measured by the R\'enyi entropy, is concerned with whether or not a state can be transformed into another state, using only local operations and classical communications. We found that in the one-dimensional Kitaev chain with quenched chemical potential $\mu$, the convertibility between the state for $\mu$ and that for $\mu+\delta\mu$, depends on the quantum phases of the system ($\delta\mu$ is a perturbation). This is similar to the adiabatic case where the ground state is considered. Specifically, when the quenched system has edge modes and the subsystem size for the partition is much larger than the correlation length of the Majorana fermions which forms the edge modes, the quenched state is locally inconvertible. We give a physical interpretation for the result, based on analyzing the interactions between the two subsystems for various partitions. Our work should help to better understand the many-body phenomena in topological systems and also the entanglement properties in the Majorana fermionic quantum computation.
\end{abstract} 

\date{\today}
\maketitle

\section{Introduction}
Topological phases of quantum matter~\cite{topo-order-1,topo-order-2}, which cannot be described by local order parameters, have been an interesting subject in condensed matter physics because of the novel properties associated with these phases, such as the topologically protected ground state degeneracy~\cite{topo-gs-degeneracy}, quantum anomalous Hall effect~\cite{QAHE-1,QAHE-2}, topological currents~\cite{topo-current} and fractional quantum Hall effect~\cite{QHE}. One prominent feature of topological phases is the possibility of creating nonlocal correlations among subsystems of the quantum matter. An ideal tool to characterize such correlations is the entanglement spectrum (ES)~\cite{ES-topo}. It is a generalization of entanglement entropy and is defined as the eigenvalues of the entanglement Hamiltonian $H_{E}$, where $H_{E}$ satisfies $e^{-H_{E}}=\rho$ ($\rho$ is the reduced state of the subsystem)~\cite{etg-H-1,etg-H-2}. Not only can ES be used to classify topological phases~\cite{ES-classify-1}
but it can also detect the zero-energy edge mode~\cite{ES-detect-edge-1,ES-detect-edge-2}. However, recent studies showed~\cite{ES-not-universal} that ES may not be universal for characterizing the quantum phases in the sense that it can exhibit singular changes within the same physical phase. Indeed, both a gapped system and a gapless one with distinct topological properties can have the same ground state~\cite{gapped-gapless-same-gs}, so that ES, when considered only for ground states, cannot distinguish between them. This also indicates that topological properties of a Hamiltonian are not only related to the ground state but also the excited states and the energy spectrum. Therefore, it might be useful to consider ES that can reflect these factors for characterizing topological phases.

On the other hand, in the quantum information community, ES (or more precisely the eigenvalues of the reduced state of the subsystem) has been used to study the local convertibility of quantum many-body states. This topic is concerned with whether or not a quantum state can be transformed into another state, using only local operations and classical communications (LOCC). Let us concentrate on the most investigated bipartite pure state $\ket{\psi_{AB}}$ of a system divided into two subsystems $A$ and $B$. As we know, under LOCC an entangled state can only be transformed into a state with the same or lower entanglement quantified e.g. by the von Neumann entanglement entropy ~\cite{Nielsen} $S_{\textrm{v}}(\rho_{A})=-\textrm{tr}(\rho_{A}\textrm{log}\rho_{A})$, where $\rho_{A}=\textrm{tr}_{B}(\ket{\psi_{AB}}\bra{\psi_{AB}})$ is the reduced state of the subsystem $A$. However, the transformed state is not arbitrary. 
The question is: which states can be attained? This was solved for bipartite pure states through considering the R\'enyi entropy~\cite{catalyst-1,catalyst-2}
\ba\label{Renyi-entropy}
S_{\alpha}(\rho_{A})=\frac{1}{1-\alpha}\textrm{log}(\textrm{Tr}\rho_{A}^{\alpha}),
\ea
where $\alpha\geq 0$. Consider two bipartite states $\ket{\psi_{AB}}$ and $\ket{\psi_{AB}'}$ with the corresponding reduced states $\rho_{A}$ and $\rho_{A}'$ respectively. If and only if $S_{\alpha}(\rho_{A})-S_{\alpha}(\rho_{A}')\geq0$ for all $\alpha\geq 0$, then $\ket{\psi_{AB}}$ can be transformed into $\ket{\psi_{AB}'}$ through LOCC, possibly with the aid of a catalyst. Here the catalyst is an entangled state that participates in the transformation process but remains intact after the transformation. The R\'enyi entropy contains all the information of the eigenvalues of $\rho_{A}$. For instance, $S_{0}(\rho_{A})=\rm{log}(d)$ ($d$ is the effective rank of $\rho_{A}$, i.e. the number of nonzero eigenvalues), $\rm{lim}_{\alpha\to1}S_{\alpha}(\rho_{A})=S_{\rm{v}}(\rho_{A})$, and $\rm{lim}_{\alpha\to\infty}S_{\alpha}(\rho_{A})=-\rm{log}\lambda_{1}$ which is the single-copy entanglement ($\lambda_{1}$ is the largest eigenvalue of $\rho_{A}$)~\cite{Orus-phdthesis}. The R\'enyi entropy has been used to study the properties of the ground state $\ket{\psi(g)}$ of the XY model in the transverse field $g$~\cite{DLC-XY}. It was shown that the differential local convertibility (DLC) of $\ket{\psi(g)}$ is related to quantum phase transitions. Here DLC refers to the transformation between $\ket{\psi(g)}$ and $\ket{\psi(g+\delta g)}$, where $\delta g$ is an adiabatic perturbation. In particular, for the transverse Ising model as a special case~\cite{DLC-Ising}, DLC is affirmative in the paramagnetic phase ($g\geq1$), while it can be negative in the ferromagnetic phase ($0\leq g<1$). It turns out that DLC is closely related to the size of the partitions and the correlation length of the system. When the correlation length increases to be comparable to the size of the subsystem, the Majorana femions (MFs), which form edge states, start recombining. As a result, DLC breaks down.

However, for other models such as one-dimensional spin-$\frac{1}{2}$ and spin-$1$ XXZ Hamiltonians, it was shown~\cite{DLC-not-for-phase} that DLC is not necessarily related to quantum phase transitions but is a good detector of explicit symmetries of the model (e.g. the SU(2) symmetry of the Heisenberg model). In other words, DLC is more closely associated with the properties of the Hamiltonian than the ground state. This result is consistent with the conclusion that ES may not be universal for characterizing the quantum phases~\cite{ES-not-universal}.

In this work, we study the local convertibility of the quantum states in the Kitaev chain~\cite{Kitaev-Majorana} with a quantum quench. The quantum quench refers to the process that the Hamiltonian is changed abruptly so that the initial ground state is no longer an eigenstate of the new Hamiltonian but evolves  with time. Considerable efforts have been made to investigate the thermalization of quenched integrable and nonintegrable systems~\cite{quench-thermal-1,quench-thermal-2,quench-thermal-add-1,quench-thermal-add-3,quench-thermal-add-5,quench-Ising-1,quench-Ising-2,quench-thermal-3,quench-thermal-4,quench-XXZ,quench-thermal-5,quench-nonintegrable-trap-1,quench-nonintegrable-trap-2,quench-nonintegrable-no-thermal}, as well as the dynamics of quenched topological edge states~\cite{quench-edge-states-1,quench-edge-states-2,quench-edge-states-3,quench-edge-states-add-1,quench-edge-states-4,quench-edge-states-5,Chung-quantum-quench,Sen-quench}. Quantum quench is essentially the natural dynamics of the system which was also studied for realizing quantum information processing in the spin systems~\cite{PST-1}.
The motivation for studying the quantum quench in our work is that the state with time evolution involves many aspects of the quenched Hamiltonian, including its excited states and the corresponding energy eigenvalues. Thus we conjecture that the local convertibility of such states may be in agreement with the quantum phases (especially the topological properties) of the physical Hamiltonian. Indeed, 
we found that our conjecture is true. The initial state for the quench is an uncorrelated state between the two subsystems. The property of DLC depends on whether or not the quenched system possesses edge modes, as well as the size of the subsystem. When the edge modes are present and the subsystem size for the partition is much larger than the correlation length of the MFs that form the edge modes, DLC is negative, otherwise DLC is affirmative. Our work points out the connection between the quantum quench and the topological properties of the system (the edge modes and MFs). Moreover, as the quench process simulates the quantum gate operations in the Majorana fermionic quantum computation (MFQC)~\cite{MF-QC}, further investigation into DLC for the quench helps to better understand the entanglement properties of the quantum states in MFQC. 

The paper is organized as follows. The Kitaev model is introduced in Sec.~\ref{Sec-2-model}, where the numerical results regarding DLC is also presented. In Sec.~\ref{Sec-3-interpret}, we interpret the results of Sec.~\ref{Sec-2-model}, based on partitioning the system into two subsystems and analyzing the interactions between them. Finally, the conclusion is given in Sec.~\ref{Sec-4-conclusion}.

\section{The model}\label{Sec-2-model}
We consider the one-dimensional Kitaev chain with the Hamiltonian~\cite{Kitaev-Majorana}
\ba\label{Kitaev-H}
H=\sum_{m}&-&w(c_{m}^{\dagger}c_{m+1}+h.c.)+\Delta(c_{m}c_{m+1}+h.c.)\nonumber\\
&-&\mu(c_{m}^{\dagger}c_{m}-\frac{1}{2}),
\ea
where $c_{m}^{\dagger}$, $c_{m}$ are the creation and annihilation operators of the electron in the $m$th site, $w$ and $\Delta$ are the nearest-neighbor hopping and pairing amplitudes, and $\mu$ is the chemical potential. The Hamiltonian~(\ref{Kitaev-H}) with periodic boundary conditions can be written in momentum space using Fourier transformation $d_{k}=\frac{1}{\sqrt{N}}\sum_{m=1}^{N}e^{-i\frac{2mk\pi}{N}}c_{m}$,
\ba\label{momentum-space}
H=-\sum_{k=-\frac{N}{2}}^{\frac{N}{2}-1}(d_{k}^{\dagger},d_{-k})[\boldsymbol{R}(k)\cdot\boldsymbol{\sigma}](d_{k},d_{-k}^{\dagger})^{T},
\ea
where $\boldsymbol{R}(k)=(0,-\Delta\sin k, w\cos k+\mu/2)$ is the pseudomagnetic field whose length times $2$ is the one-particle energy spectrum is $2R(k)=\sqrt{(2w\cos k+\mu)^{2}+4\Delta^{2}\sin ^{2}k}$, and $\boldsymbol{\sigma}=(\sigma_{x},\sigma_{y},\sigma_{z})$ is the vector of Pauli matrices. Topological properties of the Hamiltonian was discussed in Ref.~\cite{Rk-loop,Chung-quantum-quench} in terms of the winding of $\boldsymbol{R}(k)$ in the $R_{y}$-$R_{z}$ parameter space, where the region $|\mu|<2w$ is topologically distinct from $|\mu|>2w$ in the sense that the winding of $\boldsymbol{R}(k)$ for the former surrounds the origin of $R_{y}$-$R_{z}$ plane ($\Delta\neq0$), while it does not for the latter. It turns out that for $|\mu|<2w$ and $\Delta\neq0$, the Hamiltonian~(\ref{Kitaev-H}) with an infinite size and open boundary conditions supports a zero-energy edge mode composed of two unpaired Majorana fermions (a small energy splitting for the edge mode is present for a finite-size chain). The Hamiltonian~(\ref{Kitaev-H}) is also equivalent to the long wavelength limit of the modified Dirac Hamiltonian~\cite{modified-Dirac-equation} for describing the topological insulator with nontrivial $Z_{2}$ index ($=1$).

We consider a quantum quench problem as formulated below. The chain is initialized in the ground state $\ket{\psi_{0}}$ for $(w,\Delta,\mu)=(w_{0},\Delta_{0},\mu_{0})$. Then $(w,\Delta,\mu)$ are suddenly changed to $(w',\Delta',\mu')$ and the state will evolve according to the Schr\"odinger equation $i\hbar\partial_{t}\ket{\Psi(t)}=H\ket{\Psi(t)}$, i.e. $\ket{\Psi(t)}=e^{-itH}\ket{\psi_{0}}$ ($\hbar=1$). The chain can be divided into two subsystems $A$ and $B$, where $A$ ($B$) comprises $L_{A}$ ($L_{B}$) consecutive sites ($L_{A}+L_{B}=N$ is the total number of the sites in the chain). The state $\ket{\Psi(t)}$ is in general an entangled state between $A$ and $B$. If the initial state $\ket{\psi_{0}}$ is a product state between $A$ and $B$, i.e. $\ket{\psi_{0}}=\ket{\phi_{A}}\otimes\ket{\phi_{B}}$, then the entanglement generated during the time evolution can be attributed solely to the action of the Hamiltonian (\ref{Kitaev-H}). In this way, we envision that the entanglement properties of $\ket{\Psi(t)}$ will reflect some non-local properties of the Hamiltonian. The initial product state, when restricted to be the ground state of the translation invariant chain (\ref{Kitaev-H}), is either $\ket{0}_{1}\ket{0}_{2}\cdots\ket{0}_{N}$ or $\ket{1}_{1}\ket{1}_{2}\cdots\ket{1}_{N}$, where $\ket{0}_{m}$ is the vacuum state of the $m$th site ($c_{m}\ket{0}_{m}=0$ and $\ket{1}_{m}=c_{m}^{\dagger}\ket{0}_{m}$). The two states are the ground states for $\mu\to-\infty$ and $\mu\to\infty$ respectively. We also restrict ourselves to the situation that $w=\Delta$ and they are fixed during the quench. As the Hamiltonian can be scaled with $H\to H/w$, we can essentially set $w=\Delta=1$. In summary, the scenario we consider is a quantum quench where $\mu$ is suddenly changed from $\pm\infty$ to some finite value $\mu'$ while $w=\Delta=1$ is fixed. The time evolution is written as
\ba
\ket{\Psi(\mu',t)}=e^{-itH({\mu'})}\ket{\textrm{vac}},
\ea
where $\mu'$ indicates the chemical potential to be quenched, $\ket{\textrm{vac}}=\ket{0}_{1}\ket{0}_{2}\cdots\ket{0}_{N}$ is the vacuum (ground) state for $\mu\to -\infty$. In Appendix~\ref{appendix-A-proof-etg}, we shall show that the entanglement for the other case $\mu\to\infty$ is same to that for $\mu\to -\infty$.

We are interested in the asymptotic behavior of $\ket{\Psi(\mu,t)}$ as $t\to\infty$ ($\mu'$ is replaced by $\mu$ for simplicity of notation). The question is: can the state $\ket{\Psi(\mu+\delta\mu,\infty)}$ be attained from $\ket{\Psi(\mu,\infty)}$ through LOCC confined within the individual subsytems $A$ and $B$ ($\delta\mu$ is a perturbation)? If the answer is positive, the quench process with the perturbation $\delta\mu$ can be simulated by LOCC and we say that $\ket{\Psi(\mu,\infty)}$ is locally convertible (or DLC is affirmative), otherwise the process involves non-local operations between $A$ and $B$ and we say that $\ket{\Psi(\mu,\infty)}$ is locally inconvertible (or DLC is negative)~\cite{DLC-XY}. The question can be solved through considering the R\'enyi entropy in Eq.~(\ref{Renyi-entropy}). DLC is affirmative if and only if $\partial_{\mu}S_{\alpha}(\rho_{A})\geq 0$ for all $\alpha\geq 0$, or, $\partial_{\mu}S_{\alpha}(\rho_{A})\leq 0$ for all $\alpha\geq 0$. Here, it is less important which state is convertible to which. We are more concerned with the overall convertibility.

For the quadratic model (\ref{Kitaev-H}) with quantum quench, the $2^{L_{A}}$ eigenvalues of $\rho_{A}$ can be factorized as the tensor product of $L_{A}$ vectors~\cite{Quench-factorize}
\ba\label{ES-tensor-product}
\left(\begin{array}{c}
q_{1}\\
1-q_{1}
\end{array} \right)\otimes\left(\begin{array}{c}
q_{2}\\
1-q_{2}
\end{array} \right)\otimes\cdots\otimes\left(\begin{array}{c}
q_{L_{A}}\\
1-q_{L_{A}}
\end{array} \right),\nonumber\\
\ea
where $1\geq q_{1}\geq q_{2}\geq\cdots\geq q_{L_{A}}\geq1/2$ and they are the first $L_{A}$ largest eigenvalues of the $2L_{A}\times2L_{A}$ correlation function matrix (CFM): $C_{m,n}=\textrm{Tr}\rho_{A}\boldsymbol{c}_{m}\boldsymbol{c}_{n}^{\dagger}$ with $\boldsymbol{c}_{m}=(c_{m},c_{m}^{\dagger})^{T}$ and $1\leq m,n\leq L_{A}$. The CFM is obtained from its Fourier transformation \ba
G(k,t)=[1-\boldsymbol{R}(k,t)\cdot\boldsymbol{\sigma}]/2\,,
\ea
where $\boldsymbol{R}(k,t)$ is a time-dependent pseudo-magnetic field~\cite{Chung-quantum-quench} associated with the quantum quench from $\hat{\boldsymbol{R}}(k)$ to $\hat{\boldsymbol{R}'}(k)$ defined in Eq.~(\ref{momentum-space}): $\boldsymbol{R}(k,t)=\cos(4R't)\hat{\boldsymbol{R}}(k)+\sin(4R't)\hat{\boldsymbol{R}}(k)\times\hat{\boldsymbol{R}'}(k)+[1-\cos(4R't)][\hat{\boldsymbol{R}}(k)\cdot\hat{\boldsymbol{R}'}(k)]\hat{\boldsymbol{R}'}(k)$, where $\hat{\boldsymbol{R}}(k)=\boldsymbol{R}(k)/R$ and $\hat{\boldsymbol{R}'}(k)=\boldsymbol{R}'(k)/R'$. When $t\to\infty$, the oscillating terms in $\boldsymbol{R}(k,t)$ dephases and $\boldsymbol{R}(k,t)\to\boldsymbol{R}_{eff}(k)=-\hat{R}'_{z}(k)\hat{\boldsymbol{R}'}(k)$ with $\hat{\boldsymbol{R}'}(k)=(0,-\sin k,\cos k+\mu/2)/\sqrt{1+\mu\cos k+\mu^{2}/4}$, where the prime ($'$) on the right-hand side of the equation is omitted for simplicity~\cite{note-1}. The parameters $(w,\Delta,\mu)$ for the initial state are technically chosen to be $(0,0,-1)$. The R\'enyi entropy in Eq.~(\ref{Renyi-entropy}) is simplified as
\ba\label{Renyi-entropy-simplified}
S_{\alpha}(\rho_{A})=\frac{1}{1-\alpha}\sum_{j=1}^{L_{A}}\textrm{log}[q_{j}^{\alpha}+(1-q_{j})^{\alpha}],
\ea

\begin{figure}[t]
\subfigure[]{\label{Fig-1a-D10}
\includegraphics[width=3in]{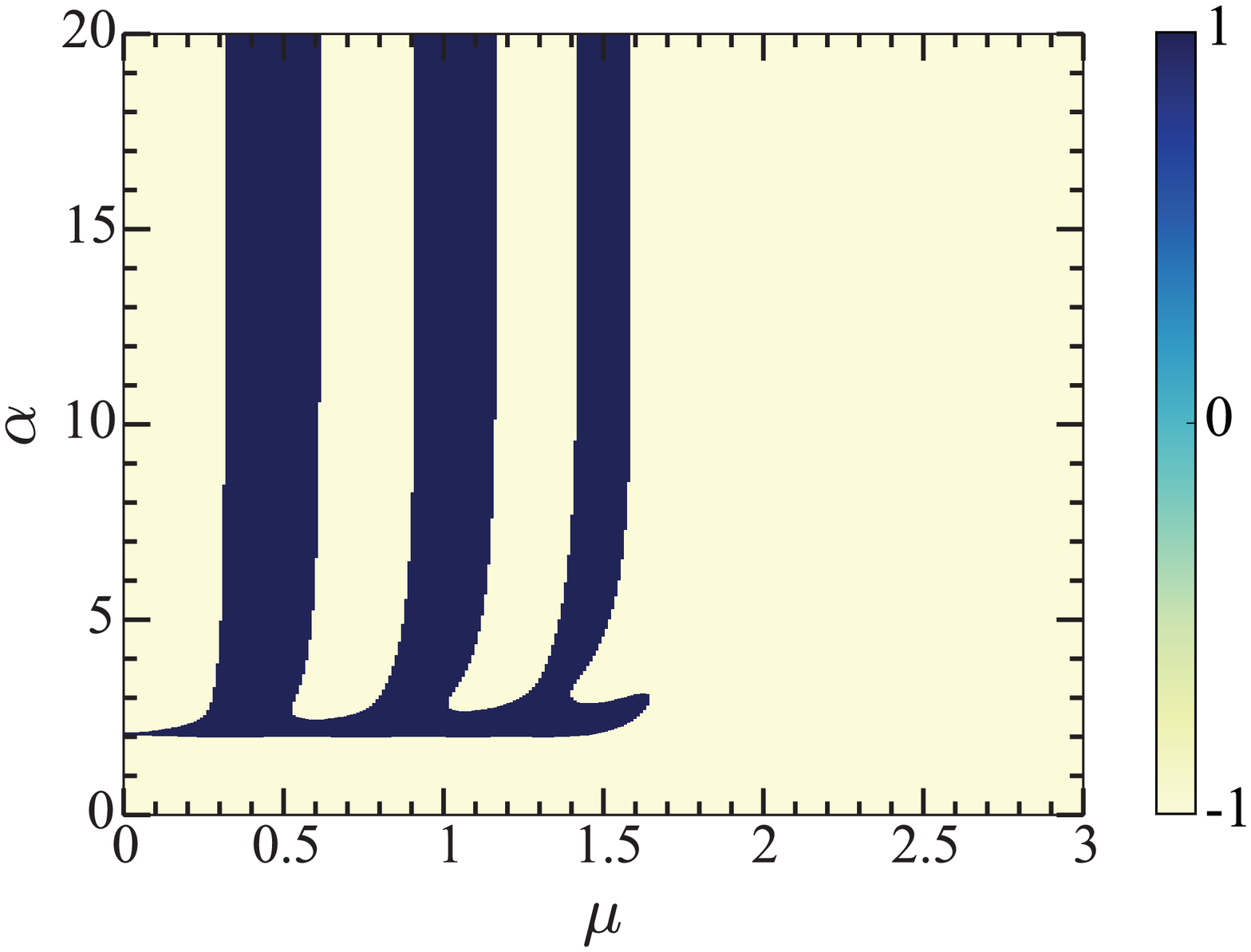}}
\vspace{3mm}
\subfigure[]{\label{Fig-1b-D50}
\includegraphics[width=3in]{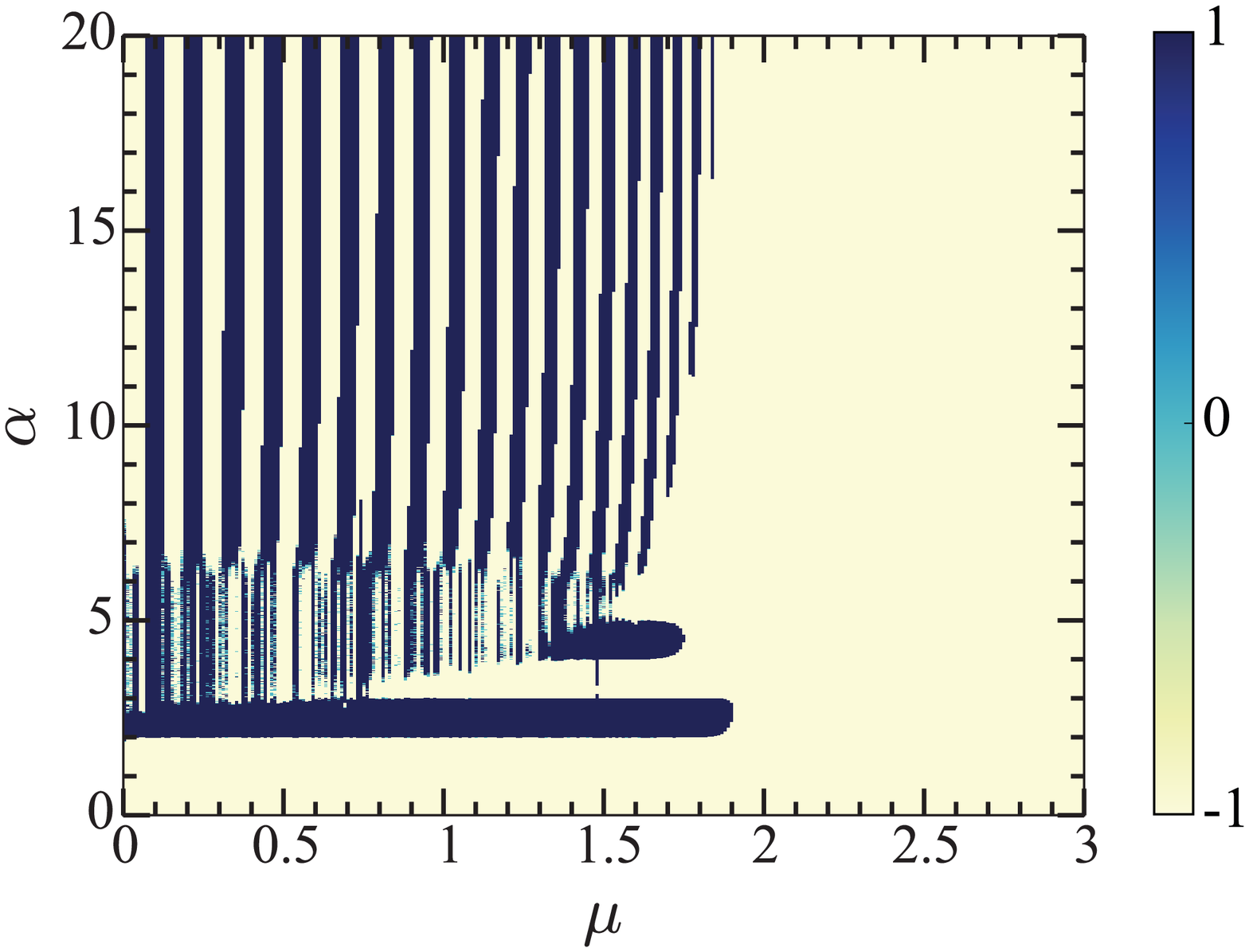}}
\caption{(Color online). The sign of the derivative of the R\'enyi entropy with respect to the quenched chemical potential $\mu$ for $0<\mu\leq3$ and $0\leq\alpha\leq20$. Here the unit of $\mu$ is the hopping amplitude $w$ which is set equal to the pairing amplitude $\Delta$ in the Hamiltonian~(\ref{Kitaev-H}). (a) The length of the subsystem $A$ is $L_{A}=10$. (b) $L_{A}=50$.}\label{Fig-1-D10-50}
\end{figure}

\begin{figure}
\includegraphics[width=0.42\textwidth,height=0.33\textwidth]{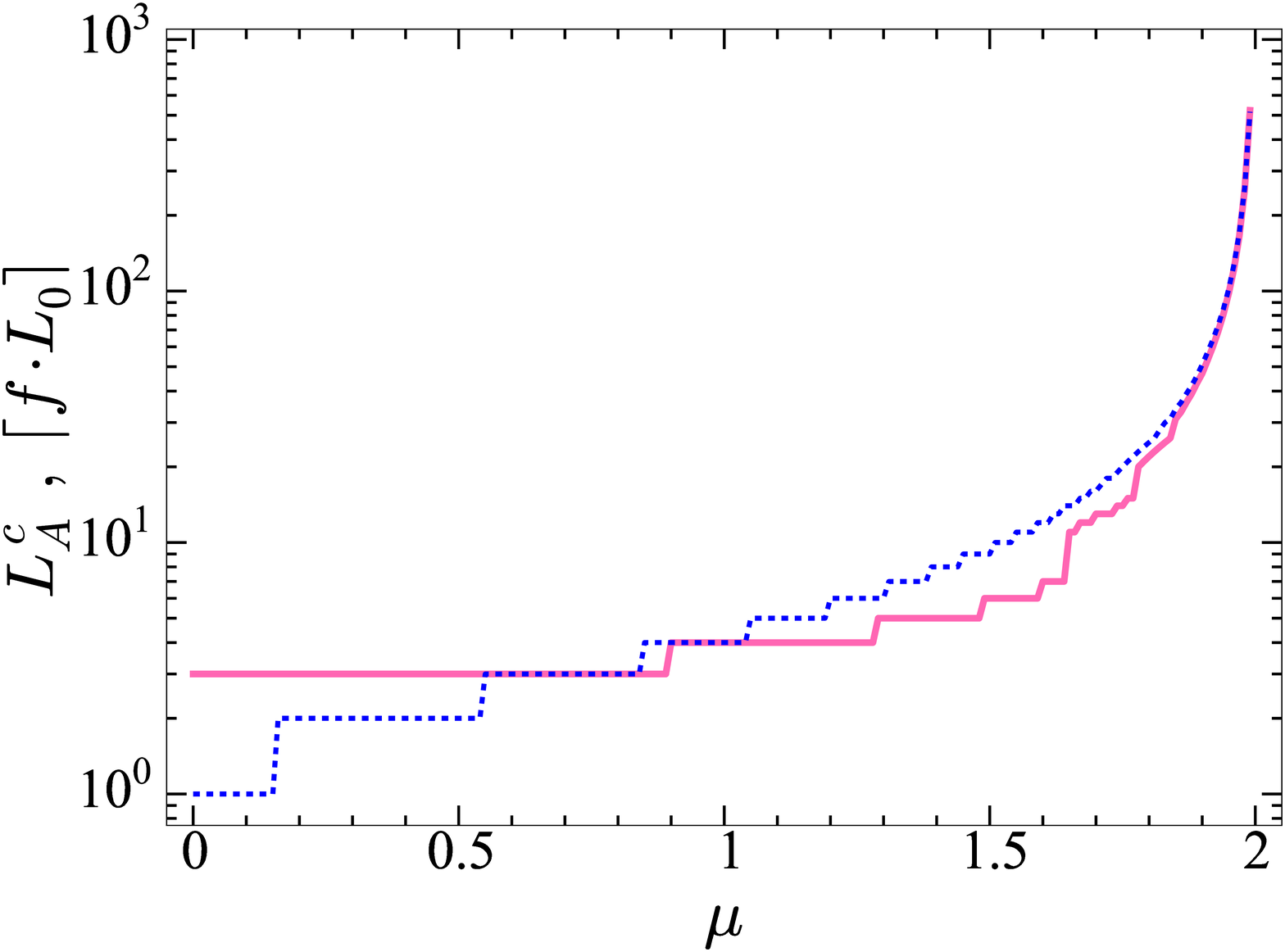}
\llap{\makebox[\wd1][l]{\raisebox{2.8cm}{\hspace{-6.2cm}\includegraphics[width=0.18\textwidth,height=0.14\textwidth]{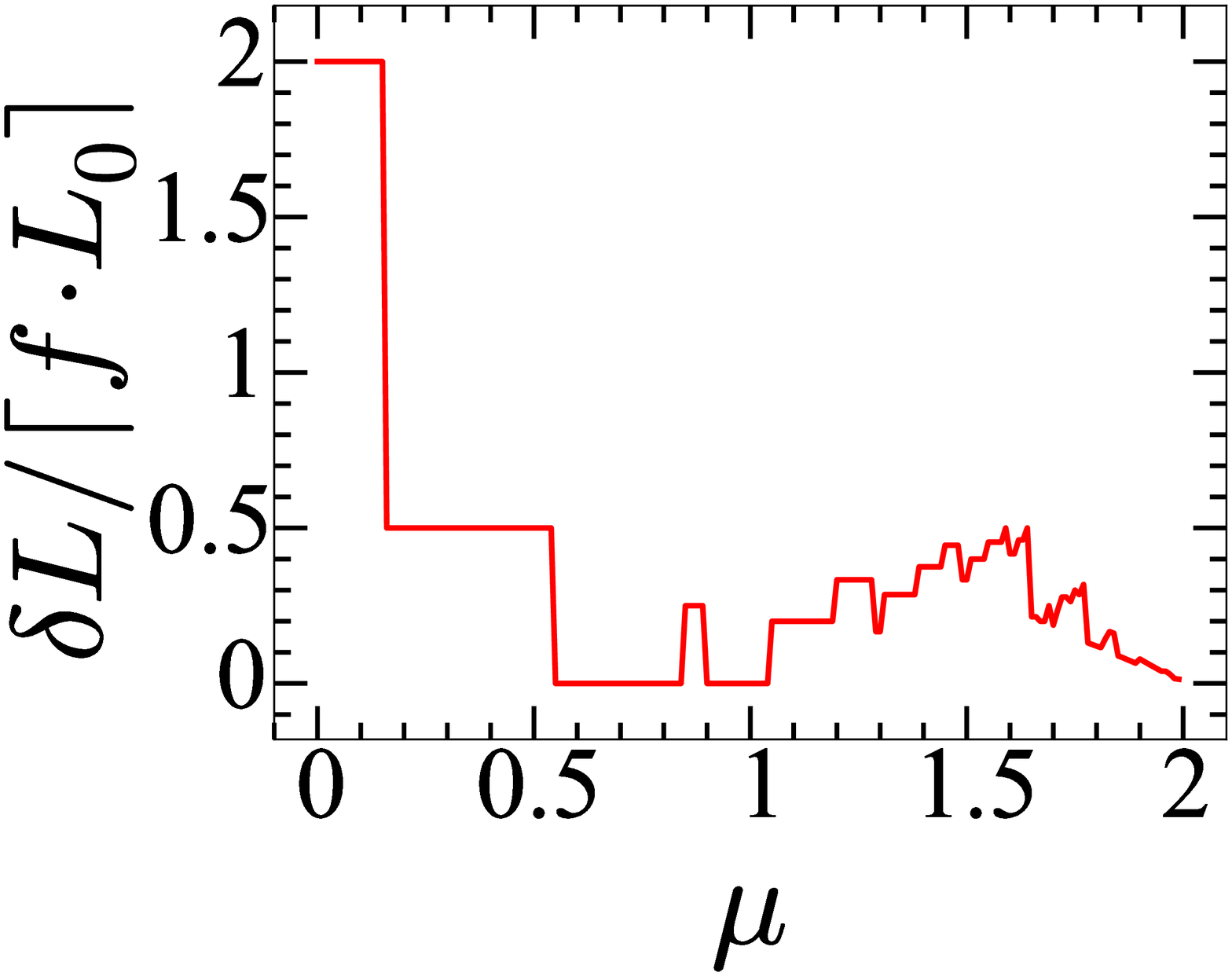}}}}
\caption{(Color online). (1) The main plot on a logarithmic scale: the solid pink line is $L_{A}^{c}$ versus $\mu$ in the unit $w(=\Delta)$; the dotted blue line is $\lceil f\cdot L_{0}\rceil$ versus $\mu$, where $\lceil x\rceil$ denotes the minimum interger lager than $x$, and $f=2.582$ is optimized for the best fitting to the solid line and $L_{0}=1/\ln(2/\mu)$. (2) The inset: the relative deviation $\delta L/\lceil f\cdot L_{0}\rceil$ versus $\mu$, where $\delta L=|L_{A}^{c}-\lceil f\cdot L_{0}\rceil|$.}\label{Fig-2-LA}
\end{figure}

We numerically analyzed DLC through calculating the R\'enyi entropy for $N\to\infty$ and different $L_{A}$'s. Fig.~\ref{Fig-1-D10-50} shows two examples with $L_{A}=10,50$. It can be seen in Fig.~\ref{Fig-1a-D10} that for $L_{A}=10$ and a fixed $\mu\in(0,1.64]$, the sign of $\partial_{\mu}S_{\alpha}(\rho_{A})$ can vary with $\alpha$, while for a fixed $\mu\in(1.64,3]$, $\partial_{\mu}S_{\alpha}(\rho_{A})$ is always negative. This result indicates that the quenched state is inconvertible for $0<\mu\leq1.64$, but it is convertible for $1.64<\mu\leq3$. Fig.~\ref{Fig-1b-D50} shows a similar result, but $1.64$ is replaced by $1.9$. Namely, the range of $\mu$ for negative DLC becomes wider. We also calculated other $L_{A}$'s and found that indeed this range widens up to $(0,2)$ as $L_{A}$ increases. Further analysis of the data shows that for $0<\mu<2$, DLC is affirmative when $L_{A}$ is smaller than some critical value $L_{A}^{c}$ as a function of $\mu$. When $L_{A}\geq L_{A}^{c}$, the results are complex: for some ranges of $\mu$ (e.g. $0<\mu\leq0.86$), DLC varies with $L_{A}$ up to a small number of increments (Max($L_{A}-L_{A}^{c})=6$), but it is always negative as $L_{A}$ increases further; for some other ranges of $\mu$ (e.g. $1.03\leq\mu\leq1.63$), DLC is always negative as long as $L_{A}\geq L_{A}^{c}$. For $\mu\geq2$, DLC is affirmative for all $L_{A}$. The behavior of DLC is reminiscent of the edge mode of the subsystem $A$. The Hamiltonian of the subsystem $A$ is Eq.~(\ref{Kitaev-H}) of $L_{A}$ sites with open boundary conditions. When $0\leq\mu<2$, there are two Majorana fermions (MFs) $\gamma_{1},\gamma_{2}$ localized at the two boundary sites of the subsystem $A$ respectively, with some overlap (a weak interaction $H_{eff}$) between them~\cite{Kitaev-Majorana}.
\ba\label{H-eff-unpaired-MFs}
H_{eff}=i\tau\gamma_{1}\gamma_{2}=\epsilon(\tilde{c}^{\dagger}\tilde{c}-\frac{1}{2}),\hspace{0.3cm}\tau\propto e^{-L_{A}/L_{0}},
\ea
where $\tilde{c}^{\dagger}=(\gamma_{1}-i\gamma_{2})/2, \tilde{c}=(\gamma_{1}+i\gamma_{2})/2$ are the creation and annihilation operators of the Dirac fermion (edge mode) formed by the two MFs, $\epsilon=2\tau$ is the energy of the Dirac fermion, and $L_{0}=1/\ln(\frac{2}{\mu})$. When $L_{A}\gg L_{0}$, $H_{eff}$ is negligible and the two MFs are said to be unpaired, with the corresponding edge mode having zero energy ($\epsilon\to0$). $L_{0}$ increases with $\mu$, and so does the overlap of the two MFs (for a fixed $L_{A}$). When $\mu$ approaches $2$, $L_{0}$ diverges ($\to\infty$) and the edge mode are absorbed into the bulk. When $\mu\geq2$, no edge mode exists. Therefore, we can set a critical length $\lceil f\cdot L_{0}\rceil$ such that an edge mode exists for $L_{A}\geq \lceil f\cdot L_{0}\rceil$, where $\lceil x\rceil$ denotes the minimum interger lager than $x$ ($L_{A}$ is an integer), and the factor $f$ designates the critical overlap of the MFs. The larger the factor $f$ is, the smaller the critical overlap of the MFs becomes. In Fig.~\ref{Fig-2-LA}, $L_{A}^{c}$ and $\lceil f\cdot L_{0}\rceil$ are plotted against $\mu$ with the optimized $f=2.582$ for the best fitting of the two curves. It can be seen that the two curves fit well. In particular, the relative deviation $\delta L/\lceil f\cdot L_{0}\rceil$ is smaller than $9\%$ when $\mu\geq 1.85$ (see the inset in Fig.~\ref{Fig-2-LA}, $\delta L=|L_{A}^{c}-\lceil f\cdot L_{0}\rceil|$). The maximum deviation between $L_{A}^{c}$ and $\lceil f\cdot L_{0}\rceil$ for $0<\mu\leq1.99$ is $7$ sites. We shall discuss their connection in detail in the next section.

\section{Physical interpretation}\label{Sec-3-interpret}
\begin{figure}[t]
\subfigure[]{\label{Fig-3a-MF-chain}
\includegraphics[width=2.5in]{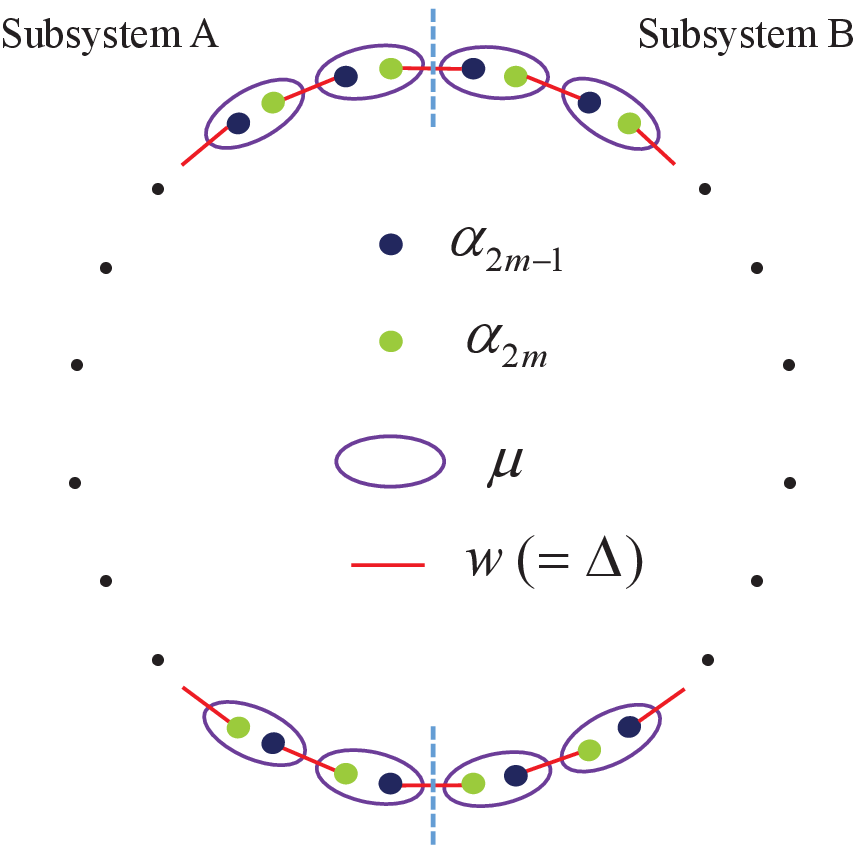}}
\vspace{3mm}
\subfigure[]{\label{Fig-3b-MF-overlap}
\includegraphics[width=2.5in]{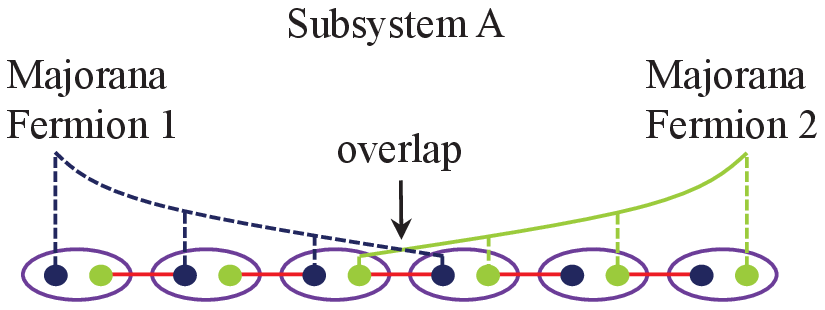}}
\caption{(Color online). (a) The Kitaev chain with $N$ sites is represented by a chain of $2N$ Majorana fermions. The vertical dashed lines mark the boundaries between the subsystems $A$ and $B$. (b) When $0\leq\mu<2$, the subsystem $A$ has an edge mode composed of two Majorana fermions (1 and 2) with a small overlap. The vertical dashed lines denote the modulus of the amplitude of the Majorana wavefunctions, and the dashed and solid curves are the overall shape of the wavefunctions. The subsystem $B$ is similar.}
\end{figure}

In this section, we shall interpret the result of the last section. To facilitate the discussion, we rewrite the Hamiltonian (\ref{Kitaev-H}) in terms of the two subsystems $A$ and $B$.
\ba\label{H-divide}
H=H_{A}+H_{B}+H_{I}\,,
\ea
where $H_{A}$ ($H_{B}$) is the Hamiltonian (\ref{Kitaev-H}) confined within the $L_{A}$ ($L_{B}$) sites of the subsystem $A$ ($B$). $H_{I}=-w(c_{L_{A}}^{\dagger}c_{L_{A}+1}+c_{N}^{\dagger}c_{1})+\Delta(c_{L_{A}}c_{L_{A}+1}+c_{N}c_{1})+h.c.$ describes the interaction between the two subsystems ($h.c.$ denotes the Hermitian conjugation of its previous terms). For $w=\Delta=1$, it is more elegant to write Eq.~(\ref{H-divide}) in the basis of Majorana operators. Define $c_{m}=(\alpha_{2m-1}+i\alpha_{2m})/2$, $c_{m}^{\dagger}=(\alpha_{2m-1}-i\alpha_{2m})/2$ for $1\leq m\leq N$, and let $\alpha_{j}\equiv\beta_{j-2L_{A}}$ for $2L_{A}+1\leq j\leq2(L_{A}+L_{B})=2N$ to indicate that these MFs belongs to the subsystem $B$. We have
\ba
\label{Kitaev-H-MF-A}H_{A}&=&i\sum_{m=1}^{L_{A}-1}\alpha_{2m}\alpha_{2m+1}-i\frac{\mu}{2}\sum_{m=1}^{L_{A}}\alpha_{2m-1}\alpha_{2m}\,,\\
\label{Kitaev-H-MF-B}H_{B}&=&i\sum_{m=1}^{L_{B}-1}\beta_{2m}\beta_{2m+1}-i\frac{\mu}{2}\sum_{m=1}^{L_{B}}\beta_{2m-1}\beta_{2m}\,,\\
\label{Kitaev-H-MF-I}H_{I}&=&i\alpha_{2L_{A}}\beta_{1}+i\beta_{2L_{B}}\alpha_{1}\,.
\ea
See Fig.~\ref{Fig-3a-MF-chain} for illustration. $H_{A}$ can be diagonalized as a sum of $L_{A}$ Dirac fermions~\cite{Kitaev-Majorana}: $H_{A}=\sum_{k=1}^{L_{A}}\epsilon_{k}^{A}(\tilde{a}^{\dagger}_{k}\tilde{a}_{k}-\frac{1}{2})$, where $\tilde{a}_{k}=(\tilde{\alpha}_{2k-1}+i\tilde{\alpha}_{2k})/2$, $\tilde{\alpha}_{j}=\sum_{m}W^{A}_{j,m}\alpha_{m}$ and $W^{A}$ block diagonalizes the coefficient matrix in $H_{A}$. $H_{B}$ is similar: $H_{B}=\sum_{k=1}^{L_{B}}\epsilon_{k}^{B}(\tilde{b}^{\dagger}_{k}\tilde{b}_{k}-\frac{1}{2})$, $\tilde{b}_{k}=(\tilde{\beta}_{2k-1}+i\tilde{\beta}_{2k})/2$, $\tilde{\beta}_{j}=\sum_{m}W^{B}_{j,m}\beta_{m}$.

We consider $N\to\infty$ but $L_{A}$ is finite (so $L_{B}\to\infty$). When $0<\mu<2$, the subsystem B supports a zero-energy edge mode, say, $\epsilon^{B}_{L_{B}}=0$, while for the subsystem $A$, only when $L_{A}\gg L_{0}$ does it support a zero-energy edge mode, say, $\epsilon^{A}_{L_{A}}\to0$, as discussed in the context below Eq.~(\ref{H-eff-unpaired-MFs}). In this situation, if we adopt the interaction picture, all the Dirac fermions oscillate except the zero-energy modes: $\tilde{a}_{k}(t)=e^{-it\epsilon^{A}_{k}}\tilde{a}_{k}$, $\tilde{b}_{k}(t)=e^{-it\epsilon^{B}_{k}}\tilde{b}_{k}$. We expect that the main contribution to the quantum state in the infinite time limit will be the steady part of the interaction Hamiltonian, which has two types. (1) For the Dirac modes with non-zero energy, when $\epsilon^{A}_{k}\approx\epsilon^{B}_{k'}$ for some $k,k'$, $H_{I}$ in Eq.~(\ref{Kitaev-H-MF-I}) in the interaction picture will have a steady part $\propto\tilde{a}_{k}^{\dagger}\tilde{b}_{k'}+h.c.$, which manifests the energy conservation. (2) For the zero-energy edge modes, the relevant steady part will be a linear combination of $\tilde{a}_{L_{A}}^{\dagger}\tilde{b}_{L_{B}}$, $\tilde{a}_{L_{A}}\tilde{b}_{L_{B}}$, and their Hermitian conjugation. In terms of Majorana fermions, this part will be
\ba\label{MF-part}
i\chi(\tilde{\alpha}_{2L_{A}}\tilde{\beta}_{2L_{B}-1}+\tilde{\beta}_{2L_{B}}\tilde{\alpha}_{2L_{A}-1}),
\ea
where $\chi=W^{A}_{2L_{A},2L_{A}}\cdot W^{B}_{2L_{B}-1,1}\,$ and the contributions relevant to $W^{A}_{2L_{A}-1,2L_{A}}$, $W^{A}_{2L_{A},1}$, $W^{B}_{2L_{B}-1,2L_{B}}$, $W^{B}_{2L_{B},1}$ is numerically found to be negligible and thus ignored. We shall argue that the second type of the steady part is the main source that renders the quenched state locally inconvertible.

For $L_{B}\to\infty$ and $0<\mu<2$, the edge mode of the subsystem $B$ can be constructed analytically~\citep{Kitaev-Majorana}, 
\ba\label{B-MF-1}
\tilde{\beta}_{2L_{B}-1}&=&\sqrt{1-\frac{\mu^2}{4}}\sum_{j=1}^{L_{B}}(-\frac{\mu}{2})^{j-1}\beta_{2j-1}\,,\\
\tilde{\beta}_{2L_{B}}&=&\sqrt{1-\frac{\mu^2}{4}}\sum_{j=1}^{L_{B}}(-\frac{\mu}{2})^{j-1}\beta_{2L_{B}-2j+2}\,.\label{B-MF-2}
\ea
The energy of the corresponding Dirac mode $\tilde{b}_{L_{B}}=(\tilde{\beta}_{1}+i\tilde{\beta}_{2L_{B}})/2$ is exactly zero. These results can be derived through examining the structure of the coefficient matrix $W^{B}$. It can be seen from Eqs.~(\ref{B-MF-1},\ref{B-MF-2}) that the two MFs are localized at the boundaries of the subsystem $B$ and their wavefunctions (the coefficient $(-\frac{\mu}{2})^{j-1}$) decay with $j$. The decay rate ($=\mu/2)$ decreases with $\mu$, indicating that the wavefunctions extend towards the inner part of the chain. The edge mode of the subsystem $A$, calculated numerically with finite $L_{A}$, has similar properties. See Fig.~\ref{Fig-3b-MF-overlap} for illustration. With the extension of the wavefunctions, the number of the sites pertaining to the edge modes increases.  We would expect that the effective dimension of the Hilbert space involved in the quench increases as well, and the entanglement becomes higher as the result of the larger dimension of the reduced state. However, this is not true, because the coefficient $\chi$ in Eq.~(\ref{MF-part}), representing the interaction strength of the MFs, decreases ($W^{B}_{2L_{B}-1,1}=\sqrt{1-\mu^2/4}$ from Eq.~(\ref{B-MF-1}) and we found numerically that $W^{A}_{2L_{A},2L_{A}}$ is well approximated by $\sqrt{1-\mu^2/4}$\,, so that $\chi\approx 1-\mu^2/4$). 

\begin{figure}[t]
\includegraphics[width=3in]{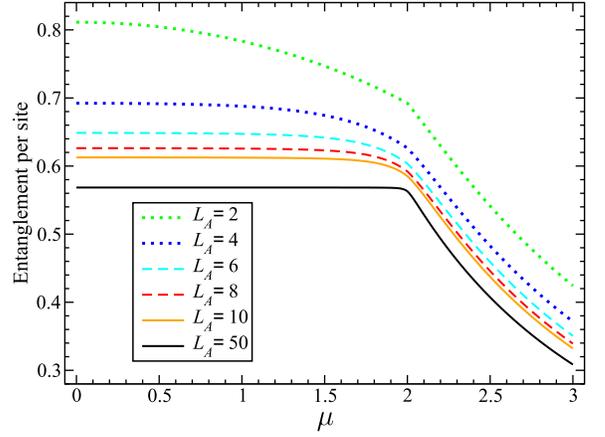}
\caption{(Color online). Entanglement (per site) of the quenched state for various length ($L_{A}$) of the subsystem $A$ is plotted against $\mu$ in the unit $w(=\Delta)$. Here the entanglement is measured by the von Neumann entropy of the reduced state of the subsystem $A$ and "per site" means that the entanglement is divided by $L_{A}$.}\label{Fig-4-etg-per-site}
\end{figure}

From the above discussions, it can be seen that there are two competing factors that affect the entanglement properties of the quenched state when $\mu$ increases: (1) The wavefunctions of the MFs within one subsystem extend inward; (2) The interaction between the MFs in different subsystems decreases. Fig.~\ref{Fig-4-etg-per-site} shows the entanglement of the quenched state versus $\mu$ for different $L_{A}$'s. At this juncture, we would like to remark that for a fixed $\mu$, the quenched state in the infinite time limit is also fixed, so that varying $L_{A}$ amounts to changing the size of the partition for a fixed state. It can be seen in Fig.~\ref{Fig-4-etg-per-site} that the entanglement decreases very slowly for small $\mu$. When $\mu$ is larger than some critical $\mu_{c}$, the entanglement starts to decrease rapidly (e.g. in Fig.~\ref{Fig-4-etg-per-site}, $\mu_{c}\sim1.6$ for $L_{A}=10$, and $\mu_{c}\sim1.9$ for $L_{A}=50$). In terms of the above two factors, we expect that when $\mu\leq\mu_{c}$, both factors exist with the second one slightly more influential. When $\mu_{c}<\mu<2$, the wavefunctions of the two MFs in the subsystem $A$ extend over the middle point of the subsystem and overlap considerably, so that the effective dimension of the Hilbert space involved in the quench saturates. Thus, the effect of the factor (1) on the entanglement becomes negligible, while the factor (2) becomes dominant, causing a rapid decrease of the entanglement. When $\mu\geq2$, the edge modes are absorbed into the bulk. It is conceivable that the entanglement decreases rapidly in this regime. This is because the bonding strength of the MFs within a single site of real space is stronger for larger $\mu$. See Fig.~\ref{Fig-3a-MF-chain}. As a result, the quenched state will be more localized to the single sites of real space, with lower entanglement.

Comparing Fig.~\ref{Fig-4-etg-per-site} with Fig.~\ref{Fig-1-D10-50} and \ref{Fig-2-LA}, we find that the value $\mu_{c}$ is consistent with the critical $\mu$ for DLC to change. This indicates that for $\mu\leq\mu_{c}$, although the entanglement, measured by the von Neumann entropy of the reduced state, decreases very slowly, the structure of the quantum state changes dramatically. The dramatic change is reflected in the fact that LOCC cannot simulate the process in which $\mu$ changes to $\mu+\delta\mu$, that is, the quenched state is locally incovertible. Some insight into the inconvertibility can be gained by examining Eq.~(\ref{ES-tensor-product}). Take $L_{A}=10$ as an example. When the quenched parameter $\mu=0.5$, the sign of $\partial_{\mu}\ln q_{j}$ roughly alternates between plus and minus when $j$ increases ($\partial_{\mu}\ln q_{1}>0$). This result influences the convertibility through the following expression derived from Eq.~(\ref{Renyi-entropy-simplified}) with $\log=\ln$.
\ba\label{cvt-du}
\partial_{\mu}S_{\alpha}(\rho_{A})=\frac{\alpha}{1-\alpha}\sum_{j=1}^{L_{A}}\frac{1-r_{j}^{\alpha-1}}{1-r_{j}^{\alpha}}\partial_{\mu}\ln q_{j},
\ea
where $r_{j}=(1-q_{j})/q_{j}$ and $0\leq r_{j}\leq1$. The sign of $\partial_{\mu}S_{\alpha}(\rho_{A})$ can vary with $\alpha$ (this is the case for $\mu=0.5$ here) on condition that the sign of $\partial_{\mu}\ln q_{j}$ is not definite when $j$ varies. Namely, the uniformity of the sign of $\partial_{\mu}\ln q_{j}$ is a sufficient but not necessary condition for the convertibility to hold. It can be verified that the former is also a sufficient condition for majorization which is stricter than the convertibility~\cite{DLC-not-for-phase}. When $\mu$ increases, the variation of the sign of $\partial_{\mu}\ln q_{j}$ with $j$ becomes less and less, while the convertibility remains to break down. When $1.64<\mu<2$, the sign of $\partial_{\mu}\ln q_{j}$ is almost uniformly plus except a few of them (no more than $3$ and $j\geq6$). In this situation, the convertibility is restored but majorization is not. When $\mu\geq2$, the sign of $\partial_{\mu}\ln q_{j}$ is plus for all $j$, which guarantees the convertibility according to Eq.~(\ref{cvt-du}) and also the majorization.

\begin{figure}[t]
\includegraphics[width=0.45\textwidth,height=0.35\textwidth]{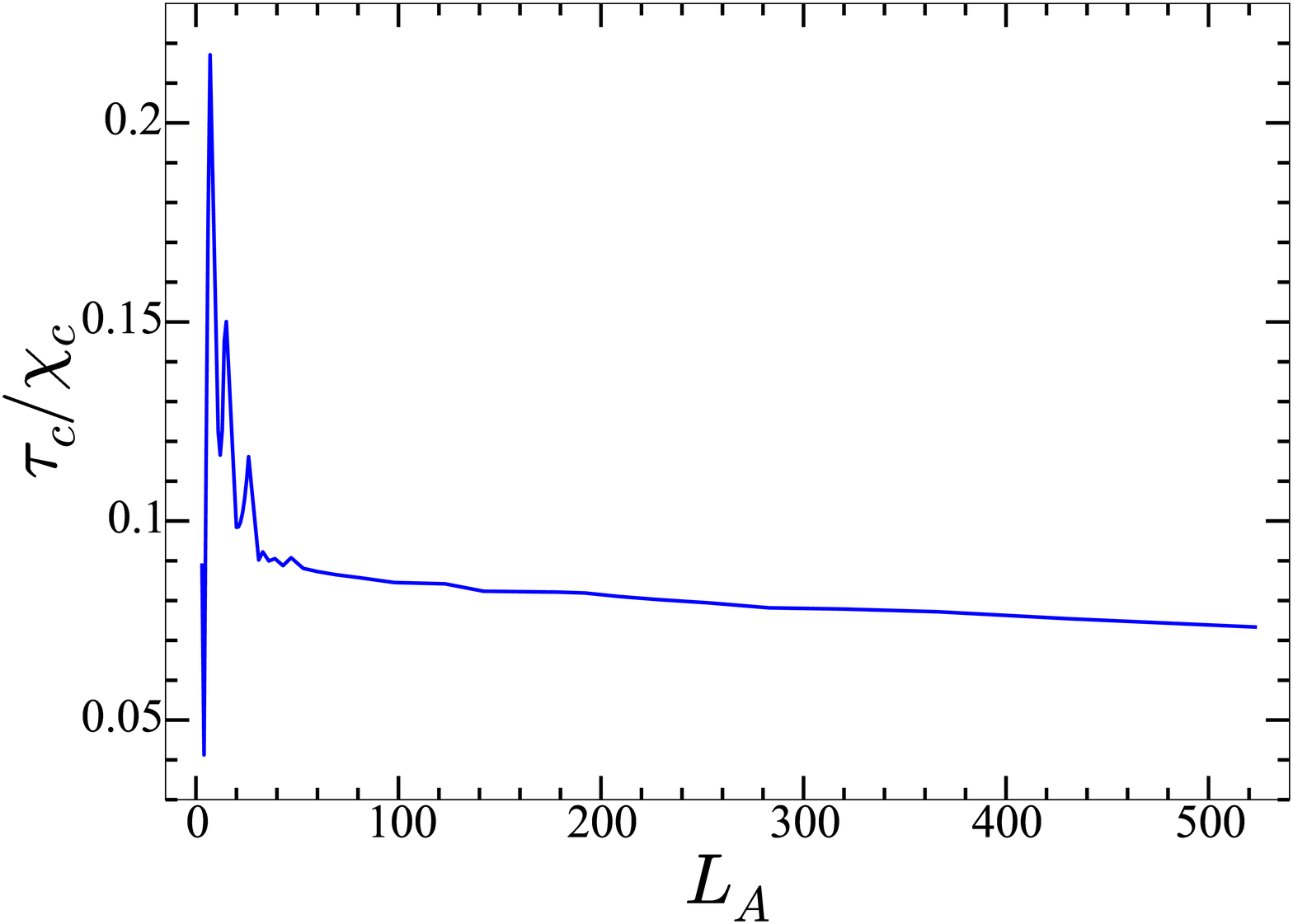}
\llap{\makebox[\wd1][l]{\raisebox{2.6cm}{\hspace{-4.3cm}\includegraphics[width=0.2\textwidth,height=0.16\textwidth]{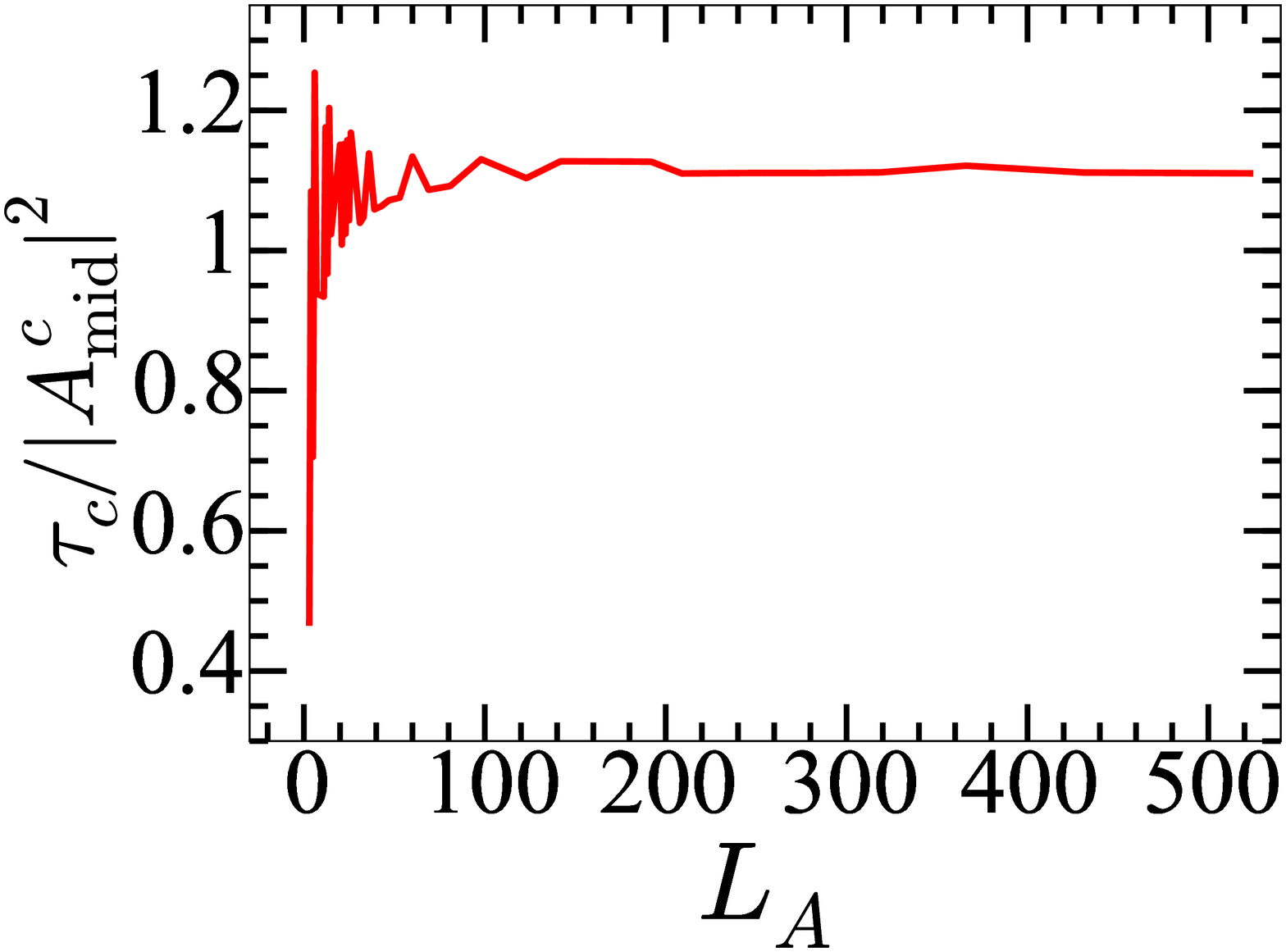}}}}
\caption{(Color online) The main plot: the ratio $\tau_{c}/\chi_{c}$ versus $L_{A}$, where $\tau_{c}$ is the critical interaction between the two Majorana fermions in the subsystem $A$, as in Eq.~(\ref{H-eff-unpaired-MFs}), and $\chi_{c}$ is the critical interaction between the Majorana fermion in the subsystem $A$ and that in the subsystem B, as in Eq.~(\ref{MF-part}). Here, ``critical" means that when $\tau/\chi>\tau_{c}/\chi_{c}$, the quenched state will change from being locally inconvertible to being locally convertible. The inset: the ratio $\tau_{c}/|A_{\rm{mid}}^{c}|^{2}$ versus $L_{A}$, where $A_{\rm{mid}}^{c}$ is the amplitude of the Majorana wavefunction in the middle point of the subsytem $A$ when $\tau/\chi=\tau_{c}/\chi_{c}$.}\label{Fig-5-ratio}
\end{figure}

To justify the analysis, the ratio $\tau_{c}/\chi_{c}$ is plotted against $L_{A}$ in Fig.~\ref{Fig-5-ratio}, where $\tau_{c},\chi_{c}$ are in Eqs.~(\ref{H-eff-unpaired-MFs},\ref{MF-part}) with the subscript ``$c$" added, and the critical $\mu,L_{A}^{c}$ in the solid curve in Fig.~\ref{Fig-2-LA} are used (the subscript ``$c$" for $\tau,\chi$ indicates ``critical"). This ratio represents the competition between the above two factors, which can be seen from the definition of $\tau$ and $\chi$. More explicitly, when $\tau/\chi>\tau_{c}/\chi_{c}$, the quenched state will change from being locally inconvertible to being locally convertible, and we expect that $\tau_{c}/\chi_{c}$ should be insusceptible to the change of $L_{A}$. Numerically, we find a slow decrease of $\tau_{c}/\chi_{c}$ for large $L_{A}$, as in Fig.~\ref{Fig-5-ratio}. For $100\leq L_{A}\leq500$, $\tau_{c}/\chi_{c}\approx 0.08\pm0.005$. This is consistent with the data fitting in Fig.~\ref{Fig-2-LA} as discussed below. As we know, $\tau_{c}$ is proportional to the overlap of the MFs. The overlap can be quantified as $|A_{1,\rm{mid}}^{c}\cdot A_{2,\rm{mid}}^{c}|=|A_{\rm{mid}}^{c}|^{2}$, where $A_{1,\rm{mid}}^{c}$\,, $A_{2,\rm{mid}}^{c}$ are the amplitudes of the two Majorana wavefunctions respectively in the middle point of the subsytem $A$ when $\tau/\chi=\tau_{c}/\chi_{c}$. Here $A_{1,\rm{mid}}^{c}=A_{2,\rm{mid}}^{c}\triangleq A_{\rm{mid}}^{c}$. In the inset of Fig.~\ref{Fig-5-ratio}, we find numerically that $\tau_{c}=1.1|A_{\rm{mid}}^{c}|^{2}$ for large $L_{A}$. Moreover, $A_{\rm{mid}}^{c}\approx \sqrt{1-\frac{\mu_{c}^{2}}{4}}(-\frac{\mu_{c}}{2})^{L_{A}^{c}/2}$ and $\chi_{c}\approx 1-\frac{\mu_{c}^{2}}{4}$ (see the context below Eq.~(\ref{B-MF-2})). Therefore, $\tau_{c}/\chi_{c}\approx 1.1(\frac{\mu_{c}}{2})^{L_{A}^{c}}=1.1e^{-L_{A}^{c}/L_{0}}|_{\mu\to\mu_{c}}=1.1e^{-2.582}=0.08$, where $L_{0}$ is given in Eq.~(\ref{H-eff-unpaired-MFs}).

\section{Conclusion}\label{Sec-4-conclusion}
We have studied the local convertibility of the quantum state in the Kitaev chain with a quantum quench, where the chemical potential of the system is suddenly changed from $\pm\infty$ to a finite value $\mu$. We found that the quenched state is locally inconvertible in the topological regime where the subsystems possess edge modes composed of Majorana fermions with weak interaction. When the interaction becomes sufficiently strong, or when no edge modes are present, the quenched state is locally convertible. However, the von Neumann entanglement entropy of the quenched state decreases for all the regimes, albeit in the incovertible situation, the rate of decrease is very small. The distinguishing behavior of the convertibility of the quenched state, as compared with the entanglement entropy, indicates that the many-body quantum state may have rich structure that cannot be well characterized by the bipartite entanglement entropy. In particular, the rich structure, characterized by the local convertibility, turns out to be closely related to the topological properties of the system (the edge modes and Majorana fermions). Thus our work should help to better understand many-body phenomena in topological systems, especially with a quantum quench.

Future work can be pursued on more general parameters, e.g. $w\neq\Delta$ in Eq.~(\ref{Kitaev-H}), in order to verify whether or not our conclusion regarding the edge modes and local convertibility is still applicable. Another interesting aspect is to examine a more general initial product state. For example, the initial state can be the tensor product of the ground states of the two subsystems. In the topological regime where edge modes are present in the subsystems, the quantum quench will involve interactions between the edge modes which amounts to performing quantum operations on the initial edge states. This is related to the Majorana fermionic quantum computation where Majorana fermions are manipulated to realize quantum gates~\cite{MF-QC}. Detailed investigation will help to understand the entanglement properties of the quantum states in the Majorana fermionic quantum computation.

\section*{Acknowledgements}
Li Dai would like to thank Yu-Chin Tzeng for helpful discussions. Li Dai is supported by the MOST Grant under the Contract Number: 103-2811-M-005-013 in Taiwan. M.-C. Chung is supported by the MOST Grant under the Contract Number: 102-2112-M-005-001-MY3.

\appendix
\section{Identical Entanglement for quench from $\mu\to\pm\infty$ to $(w,\Delta,\pm\mu)$}\label{appendix-A-proof-etg}
We shall prove that the quantum states with the following four quenches have the same entanglement, i.e. the eigenvalues of the reduced states are identical.
\ba
e^{-itH(w,\Delta,\pm\mu)}\ket{\textrm{ful}},\hspace{0.3cm}e^{-itH(w,\Delta,\pm\mu)}\ket{\textrm{vac}},
\ea
where $\ket{\textrm{vac}}=\ket{0}_{1}\ket{0}_{2}\cdots\ket{0}_{N}$, $\ket{\textrm{ful}}=\ket{1}_{1}\ket{1}_{2}\cdots\ket{1}_{N}$ ($\ket{0}_{m}$ is the vacuum state of the $m$th site: $c_{m}\ket{0}_{m}=0$, and $\ket{1}_{m}=c_{m}^{\dagger}\ket{0}_{m}$).

Let us consider the equivalence between $e^{-itH(w,\Delta,\mu)}\ket{\textrm{ful}}$ and $e^{-itH(w,\Delta,\mu)}\ket{\textrm{vac}}$. The Fourier transformed correlation function matrix (CFM) for the former is $G_{1}(k,t)=[1-\boldsymbol{R}_{1}(k,t)\cdot\boldsymbol{\sigma}]/2$, while for the latter it is $G_{2}(k,t)=[1-\boldsymbol{R}_{2}(k,t)\cdot\boldsymbol{\sigma}]/2$. Here, $\boldsymbol{R}_{1}(k,t)=-\boldsymbol{R}_{2}(k,t)\equiv\boldsymbol{R}(k,t)$ whose detailed formula is presented in Ref.~\cite{Chung-quantum-quench}. The eigenvalues of the reduced state of the subsystem $A$ are the first $L_{A}$ largest eigenvalues of the CFM in real space $C_{m,n}=\textrm{Tr}\rho_{A}\boldsymbol{c}_{m}\boldsymbol{c}_{n}^{\dagger}$ with $\boldsymbol{c}_{m}=(c_{m},c_{m}^{\dagger})^{T}$. As $G_{1}(k,t)$ and $G_{2}(k,t)$ only differ on a minus sign, if the eigenvalue of $C_{m,n}$ for $G_{1}(k,t)$ is $(1+\lambda)/2$, ($0\leq\lambda\leq1$), there must be an eigenvalue $(1-\lambda)/2$ for $G_{2}(k,t)$. Moreover, the eigenvalues of $C_{m,n}$ can be written in pairs $(1\pm\lambda)/2$~\cite{Quench-factorize}. We conclude that the eigenvalues of $C_{m,n}$ for $G_{1}(k,t)$ and $G_{2}(k,t)$ are same. This completes the proof.

Next, we prove that $e^{-itH(w,\Delta,\mu)}\ket{\textrm{vac}}$ has the same entanglement as $e^{-itH(w,\Delta,-\mu)}\ket{\textrm{vac}}$. As discussed above, the Fourier transformed CFM for the former is $G_{2}(k,t)=[1-\boldsymbol{R}_{2}(k,t)\cdot\boldsymbol{\sigma}]/2$, while for the latter, it is $G_{3}(k,t)=[1-\boldsymbol{R}_{3}(k,t)\cdot\boldsymbol{\sigma}]/2$, where $\boldsymbol{R}_{3}(k,t)$ corresponds to a quench process to $(w,\Delta,-\mu)$. We notice that $C_{m,n}$ for $G_{3}(k,t)$ is $C_{m,n}=\frac{1}{2\pi}\int_{-\pi}^{\pi}G_{3}(k,t)e^{ik(m-n)}dk$. If the integration variable $k$ is shifted by $\pi$, i.e. $k\to k'=k-\pi$, we obtain an extra phase $e^{i\pi(m-n)}$ in the integrand. If this phase is negelected, the eigenvalues of $C_{m,n}$ won't change (this can be found by examining the entries of $C_{m,n}$). Therefore, we essentially have $G_{3}(k,t)\to G_{3}(k+\pi,t)=[1-\boldsymbol{R}_{3}(k+\pi,t)\cdot\boldsymbol{\sigma}]/2$. Furthermore, $\boldsymbol{R}_{3}(k+\pi,t)=-\boldsymbol{R}_{2}(k,t)$. Thus, following the same argument in the previous paragraph, we conclude that the eigenvalues of the reduced states for $e^{-itH(w,\Delta,\mu)}\ket{\textrm{vac}}$ and $e^{-itH(w,\Delta,-\mu)}\ket{\textrm{vac}}$ are same. The proof for the equivalence between the entanglement of $e^{-itH(w,\Delta,\mu)}\ket{\textrm{ful}}$ and that of $e^{-itH(w,\Delta,-\mu)}\ket{\textrm{ful}}$ is similar and omitted.

\bibliography{apssamp}

\begin{thebibliography}{53}
\bibitem [{not()}]{note-1}%
  \BibitemOpen
  \href@noop {} {}\bibinfo {note} {We have numerically verified that for
  $\mu\neq0$ the time evolution of the reduced-state eigenvalues converges to
  those associated with $\boldsymbol{R}_{eff}(k)$. For $\mu=0$, the
  Hamiltonian~(\ref{Kitaev-H}) is equivalent, through using the reversed
  Jordan-Wigner transformation~\cite{DLC-Ising}, to the Ising-type interaction
  and the quench essentially creates the one-dimensional cluster
  state~\cite{cluster-state} at the special time $t=j\pi/(4w)$, where $j$ is an
  odd integer. Therefore, the reduced-state eigenvalues will oscillate
  periodically with time and no stationary state can be reached. Although this
  case is interesting in the quantum information
  science~\cite{cluster-state,cluster-QC}, we shall focus on the stationary
  situation ($\mu\neq0$) in the present work.}\BibitemShut {Stop}%
\end{thebibliography}



\end{document}